\documentclass[10pt]{article}
\usepackage{graphicx}
\usepackage{url}
\usepackage{subfigure}
\usepackage{amsmath}

\title{Universality of scholarly impact metrics}

\author{Jasleen Kaur, Filippo Radicchi, Filippo Menczer \\
\\
\normalsize{Center for Complex Networks and Systems Research}\\
\normalsize{School of Informatics and Computing, Indiana University, Bloomington, USA}\\
}

\date{}
\begin{document}

\maketitle
    
  \begin{abstract}
%% Text of abstract
Given the growing use of impact metrics in the evaluation of scholars, journals, academic institutions, and even countries, there is a critical need for means to compare scientific impact across disciplinary boundaries. Unfortunately, citation-based metrics are strongly biased by diverse field sizes and publication and citation practices. As a result, we have witnessed an explosion in the number of newly proposed metrics that claim to be ``universal.'' However, there is currently no way to objectively assess whether a normalized metric can actually compensate for disciplinary bias. We introduce a new method to assess the universality of any scholarly impact metric, and apply it to evaluate a number of established metrics. We also define a very simple new metric $h_s$, which proves to be universal, thus  allowing to compare the impact of scholars across scientific disciplines. These results move us closer to a formal methodology in the measure of scholarly impact.
\end{abstract}

%\begin{keyword}
%%% keywords here, in the form: keyword \sep keyword
%impact metrics \sep discipline bias \sep universality \sep bibliometrics 
%\end{keyword}

%\end{frontmatter}

%% \linenumbers

%% main text

%Introduction 
%Material and methods 
%Theory/calculation 
%Results 
%Discussion 
%Conclusions 
%Appendices 

\section{Introduction}

Objective evaluation of scientific production --- its quantity, quality, and impact --- is quickly becoming one of the central challenges of science policy with the proliferation of academic publications and diversification of publishing outlets~\cite{abbott2010metrics}. 
Many impact metrics have been and continue to be proposed~\cite{van2010metrics}, 
most of them based on increasingly sophisticated citation analysis~\cite{lane2010let}. 
These metrics have found wide applicability in the evaluation of scholars, journals, institutions, and countries~\cite{garfield2006history,davis1984faculty,kinney2007national,king2004scientific}.
%and even in the prediction of future impact~\cite{acuna2012future}. 
Unfortunately, there is very little work on quantitative assessment of the effectiveness of these metrics~\cite{bollen2009principal,radicchi2012testing} and the few existing efforts are proving highly controversial~\cite{lehmann2006measures}. This is alarming, given the increasingly crucial role of impact analysis in grant evaluation, hiring, and tenure decisions~\cite{bornmann2006selecting}. 
%\cite{bornmann2006selecting, bornmann2008does}. 

%identify the problem, also emphasize on inter-disciplinary research

\begin{figure}[t!]
\centerline{\includegraphics[width=\textwidth]{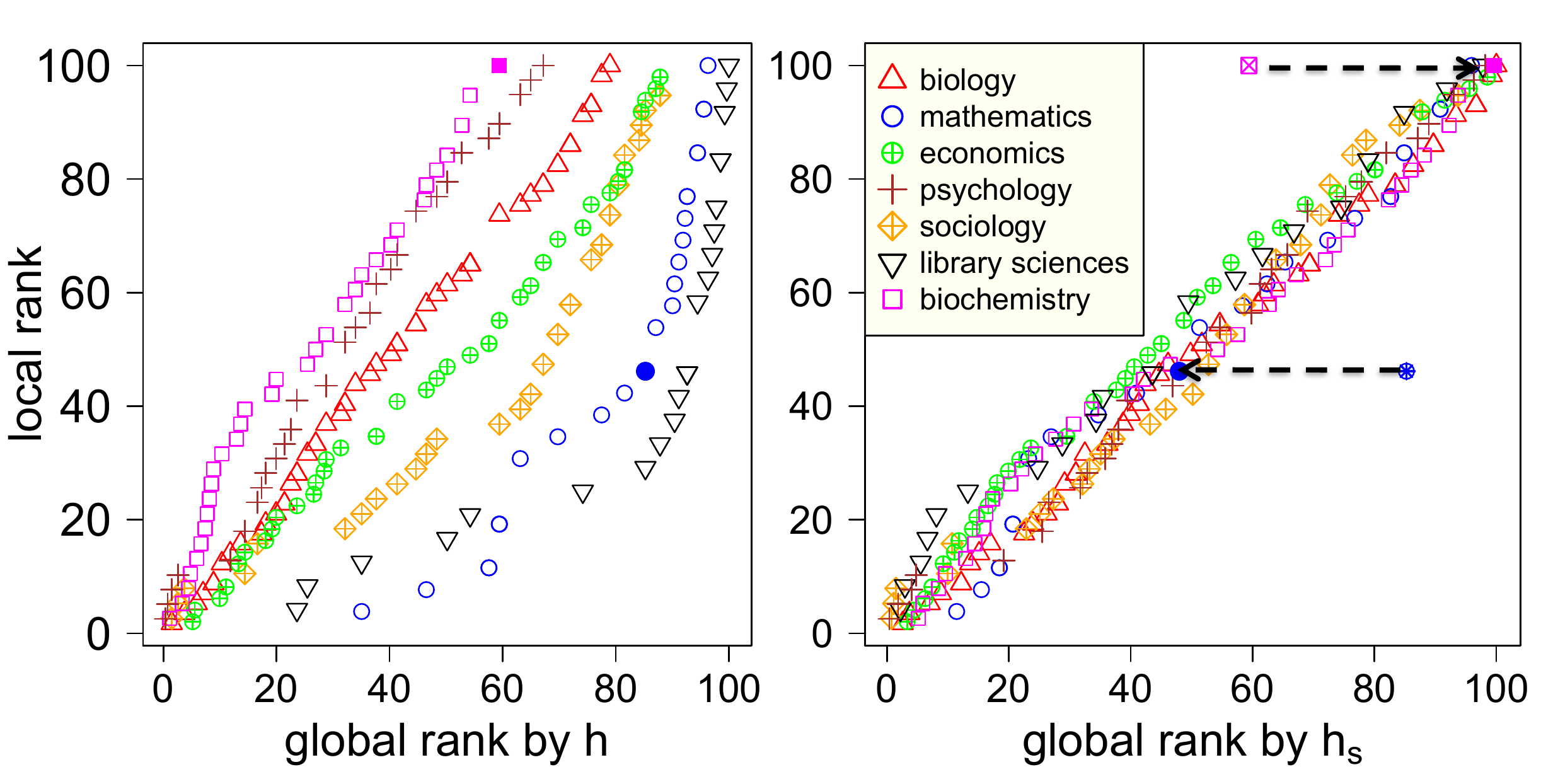}}
\caption{Effect of field normalization on ranking bias. We rank the top 5\% of authors in seven JCR disciplines according to two metrics, $h$ and $h_s$ (see text). We compare the rank (top percentile) globally across disciplines versus locally within an author's own discipline. Due to discipline bias, biochemists are favored and mathematicians are penalized according to $h$, as illustrated by the two highlighted authors. The global ranking according to the normalized metric $h_s$ is more consistent with the rankings within disciplines.}
\label{fig:compare_a}
\end{figure}

Discipline bias is probably the most critical and debated issue in impact metric evaluation. Publication and citation patterns vary wildly across disciplines, due to differences in breadth and practices. These differences introduce strong biases in impact measures --- a top scholar in biology has a very different publication and citation profile than one in mathematics. This has led to a recent burst of interest in \emph{field normalization} of impact metrics, and the emergence of many ``universal" metrics that claim to compensate for discipline bias~\cite{canadareport}. 
Fig.~\ref{fig:compare_a} illustrates the idea of field normalization. If we rank scholars across all disciplines according to an unbiased (universal) metric, a scholar in the top 5\% among mathematicians should be ranked the same as a scholar in the top 5\% among biochemists. A biased metric on the other hand may favor some disciplines and penalize others.

%~\cite{canadareport, moed2004handbook}. 

\section{Methods}

\subsection{Data}

We used the data collected by Scholarometer (\url{scholarometer.indiana.edu}) from November 2009 to August 2012.
Scholarometer is a social tool for scholarly services developed at Indiana University, with the goal of exploring the crowdsourcing approach for disciplinary annotations and cross-disciplinary impact metrics~\cite{hoang2010crowdsourcing,Kaur2012}. Users provide discipline annotations (tags) for queried authors, which in turn are used to compare author impact across disciplinary boundaries. The data collected by Scholarometer is available via an open API. We use this data to compute several impact metrics for authors belonging to various disciplines, and test the universality of these metrics. As of August 2012, the database had collected citation data about 38 thousand authors of 2.2 million articles in 1,300 disciplines.
Further statistics for authors and disciplines are available on the Scholarometer website~\cite{Kaur2012}.

%\pagebreak

\subsection{Impact metrics}
\label{impact}

The bibliometrics literature contains a plethora of scholarly impact metrics, and it is not feasible to evaluate all of them. Therefore we focus on a small set of metrics that are widely adopted and/or specifically designed to mitigate discipline bias. Our analysis of universality is performed on the following impact metrics:

\begin{description}

\item[$c_{avg}$] is the average number of citations received by an author's articles.

\item[$h$ index] is defined as the maximum number of articles $h$ such that each has received at least $h$ citations~\cite{Hirsch2005}. The $h$ index is the most widely adopted impact metric. It summarizes the impact of a scholar's career using a single number without any threshold. 

\item[Redner's index $c_{total}^{1/2}$] is defined as the square root of the total number of citations received by an author's articles~\cite{redner2010meaning}.

\item[$h_m$ index]  attempts to apportion citations fairly for papers with multiple authors~\cite{Schreiber2008}. It counts the papers fractionally according to the number of authors. This yields an effective rank, which is utilized to define $h_m$ as the maximum effective number of papers that have been cited $h_m$ or more times. 

\item[$g$ index] is the highest number $g$ of papers that together receive $g^2$ or more citations~\cite{LeoEgghe2006}. It attempts to mitigate the insensitivity of the $h$ index to the number of citations received by highly cited papers.

\item[$i_{10}$] is proposed by Google and is defined as the number of articles with at least ten citations each~\cite{googlei10}.

\item[$h_f$ index] was proposed as a universal variant of $h$~\cite{Radicchi2008}. The number of citations $c$ received by each paper is normalized by the average number of citations $c_0$ for papers published in the same year and discipline. The rank of each paper $n$ is rescaled by the average number $n_0$ of papers per author written in the same year and discipline. The $h_f$ index of the author is the maximum rescaled rank $h_f$ such that each of the top $h_f$ papers has at least $h_f$ rescaled citations. 

\item[Batista's $h_{i,norm}$] involves normalizing the total number of citations in the $h$-core (the papers that contribute to the $h$ index) by the total number of authors contributing to them. The resulting $h_i$ of each author is then normalized by the average $h_i$ of the author's discipline~\cite{batista2006possible}.

\item[New crown indicator $(c/c_0)_{avg}$]  was proposed by Lundberg~\cite{lundberg2007lifting} as the item oriented field-nor\-malized citation score (FNCS) and implemented by Waltman \textit{et al.}~\cite{waltman2011towards}. It is calculated as the average field-normalized number of citations $c/c_0$ across an author's publications.

\item[$h_s$ index] is proposed here as a normalization of the $h$ index by the average $h$ of the authors in the same discipline. Numerical tests show that the distribution of $h$ is not scale-free and therefore the mean is well defined. Despite its simplicity, we are not aware of this metric being previously defined in the literature. Note that within a discipline, $h_s$ produces the same ranking as $h$. Therefore, $h_s$ is very similar to the percentile score but slightly easier to compute. Percentiles have been proposed for normalization of journal impact factors~\cite{leydesdorff2011integrated}. 

\end{description}

\subsection{Disciplines}
\label{disciplines}

To test the universality of the different impact metrics, we consider three distinct ways to define disciplines, i.e., to sample authors from multiple disciplines. When a user queries the Scholarometer system, she has to annotate the queried author with at least one discipline tag from the JCR science, social sciences, or arts \& humanities indices. Additionally, the user may tag the author with any number of arbitrary (JCR or user-defined) discipline labels. Based on these annotations, we consider three disciplinary groupings of authors:

\begin{description}
\item[ISI:] The 12 JCR disciplines with the most authors (see Table~\ref{ISITop10}).
\item[User:] The top 10 user-defined disciplines (Table~\ref{UserTop10}). 
\item[Manual:] 11 manually constructed groups of related disciplines (Table~\ref{ManualTop10}).
\end{description}

\begin{table}
\caption{Top JCR (ISI) disciplines. In this and the following tables, we display the average $h$ index computed across authors in the same discipline.}
\begin{center}
\begin{tabular}{rlrr}
\hline
& Discipline & Authors & $\langle h \rangle$\\
\hline
1. & computer science, artificial intelligence & 1,922 & 15.96\\
2. & biology & 1,147 & 19.66\\
3. & economics & 972 & 17.02\\
4. & engineering, electrical \& electronic & 936 & 14.77\\
5. & neurosciences & 840 & 22.95\\
6. & political science & 794 & 15.81\\
7. & psychology & 774 & 21.18\\
8. & biochemistry \& molecular biology & 766 & 22.37\\
9. & sociology & 749 & 16.70\\
10. & mathematics & 516 & 13.55\\
11. & philosophy & 501 & 13.63\\
12. & information science \& library science & 480 & 11.15\\
\hline
\end{tabular}
\end{center}
\label{ISITop10}
\end{table}
 
 \begin{table}
\caption{Top user-defined disciplines.}
\begin{center}
\begin{tabular}{rlrr}
\hline 
& Discipline & Authors & $\langle h \rangle$\\
\hline
1.& computer science & 656 & 16.02\\
2.& physics & 200 & 18.66\\
3.& computer networks & 130 & 16.25\\
4.& bioinformatics & 125 & 16.50\\
5.& engineering & 115 & 11.46\\
6.& medicine & 104 & 23.47\\
7.& chemistry & 103 & 13.92\\
8.& human computer interaction & 94 & 17.72\\
9.& computer science, security & 82 & 19.32\\
10. & image processing & 80 &18.39\\
\hline
\end{tabular}
\end{center}
\label{UserTop10}
\end{table}

\begin{table}
\caption{Manually clustered disciplines.}
{\scriptsize
\begin{center}
\begin{tabular}{rp{3cm}p{6cm}rr}
\hline 
& Manual label & Disciplines & Authors & $\langle h \rangle$\\
\hline
1. & computer science & computer science, artificial intelligence \newline image processing \newline computer networks \newline computer science \newline computer science, theory \& methods \newline  computer science, software engineering \newline computer science, information systems \newline computer science, hardware \& architecture \newline computer science, cybernetics & 4,342 & 15.79\\
2. & biology & plant sciences biology \newline zoology \newline plant sciences \newline evolutionary biology \newline entomology \newline biology \newline biodiversity conservation \newline biochemistry \& molecular biology & 2,385 & 19.56 \\
3. & behavioral sciences & sociology \newline psychology, social \newline psychology, applied \newline anthropology \newline psychology \newline behavioral sciences & 1,846 & 17.97\\
4. & engineering & engineering, mechanical \newline engineering, electrical \& electronic \newline engineering, biomedical & 1,302 & 14.93\\
5. & economics & economics & 972 & 17.02\\
6. & mathematics & statistics \& probability \newline mathematics, applied \newline mathematics & 860 & 15.53\\
7. & political science & public administration \newline political science & 812 & 15.74\\
8. & physics & physics, applied \newline physics, multidisciplinary \newline physics, condensed matter \newline physics & 675 & 19.63\\
9. & business & business, marketing \newline management \newline business, finance \newline business & 665 & 15.59\\
10. & education \& educational research & education technology \newline education \& educational research & 305 & 12.18 \\
11. & humanities, multidisciplinary & humanities, multidisciplinary \newline humanities & 122 & 9.00\\
\hline
\end{tabular}
\end{center}}
\label{ManualTop10}
\end{table}

In Section~\ref{results}, we present results based on the ISI classification. In Section~\ref{discipline_sensitivity}, we analyze the robustness of our results against the three disciplinary groupings of authors.

\section{Theory}
\label{theory}

An objective, quantitative assessment of metric universality is missing to date. To fill this void, we introduce a \emph{universality index} to evaluate and compare the bias of different metrics.  Our index allows for the first time to gauge a metric's capability to compare the impact of scholars across disciplinary boundaries, creating an opportunity for, say, mathematicians and biologists to be evaluated consistently.

%from Radicchi and Castellano paper
The proposed universality index looks at how top authors according to a particular metric are allocated across different disciplines, and compares this distribution with one obtained from a random sampling process. This approach is inspired by a method for comparing expected and observed proportions of top cited papers to evaluate normalized citation counts~\cite{Radicchi2008}.  The idea is that each discipline should be equally represented in a sample of top authors. For example, if we rank scholars across all disciplines according to an unbiased (universal) metric, the top 5\% of scholars should include the top 5\% of mathematicians, the top 5\% of biologists, and so on. In other words, the \emph{percentage} of top scholars in each discipline should not depend on the size of the discipline. Of course the \emph{number} of scholars in each discipline should be proportional to the size of that discipline.

Suppose each author is assigned a discipline $d$. For simplicity, let us assume that each author belongs to only one category. Selecting a fraction $z$ of top scholars from the entire set according to a universal metric should be equivalent to sampling a fraction $z$ of scholars at random. If we define $f_{z,d}$ as the fraction of authors belonging to discipline $d$ in the top $z$, the expected fraction is $E[f_{z,d}]= z$. 

%, the actual fractions $\vec{f_z}=\{f_z^d\}$ from all disciplines fluctuate according to a multivariate hypergeometric distribution~\cite{Radicchi2012F}. 
%
%We simulate this unbiased selection process numerically to approximate the distribution of the proportions $\vec{f_{z}^r}$ of authors present in a random sample $r$. 

%\subsection{Null model}

To understand the fluctuations in the numbers of authors from each category, consider a set of $N$ authors in $D$ categories.  Let $N_d$ be the number of authors in category $d$. Each author has a score calculated according to the rules of the particular indicator we want to test. Imagine extracting the top fraction $z$ of authors according to their scores. This list has $n_{z} = \lfloor z N \rfloor$ authors. % ($\lfloor x \rfloor$ stands for the largest integer number smaller than or equal to x). 
If the numerical indicator is fair, the selection of an author in category $d$ should depend only on the category size $N_d$, and not on other features that may favor or hinder that particular category. Under these conditions, the number of authors $n^z_d$ in category $d$ that are part of the top $z$ is a random variate obeying the hypergeometric distribution~\cite{Radicchi2012F}:
\begin{equation}
P\left(n^z_d \left| n_{z},N,N_d\right.\right) = \binom{N_d}{n^z_d} \binom{N-N_d}{n_{z}-n^z_d}\left/\binom{N}{n_{z}}\right.
\label{mahmoud}
\end{equation}
where $\binom {x}{y}=\frac{y!}{x!(x-y)!}$ is a binomial coefficient that calculates the total number of ways in which $y$ elements can be extracted out of $x$ total elements. Eq.~\ref{mahmoud} describes a simple urn model~\cite{mahmoud2008}, where elements (authors in our case) are randomly extracted from the urn without replacement. Such a random process provides us with a  \emph{null model} for the values of $f_{z,d}$.

%ADDED ABOVE AS PER SUGGESTION OF REVIEWER 1 and FR
%$f_{z,d}$ (i.e., the fraction of scientists of discipline $d$ in the top $z\%$

%
%With this statistical model we can calculate the expected number of authors in category $d$ present in the top fraction $z$ as $E\left(m^z_d\right)=n_{z} N_d/N$. 
%
In Section~\ref{results} we estimate confidence intervals by simulating $10^3$ times the null model leading to Eq.~\ref{mahmoud}.  
%  to obtain the grey areas in Figs.~\ref{fig:combined_b}, \ref{fig:box_c}, \ref{fig:universality_d}, \ref{fig:confidence_ISI} and \ref{fig:All}.

%Such a random process provides us with a  null model %term of comparison
%for the values of $f_{z,1}^m, f_{z,2}^m, \ldots, f_{z,D}^m$
%%$\vec{f_{z}^m}$ 
%observed in the empirical data for all $D$ disciplines and 
%calculated according to a particular metric $m$. 

To obtain a quantitative criterion for the universality of $m$ with respect to a set of $D$ disciplines and a fraction $z$ of top scholars, we compute the \emph{universality} of metric $m$ as 
%$u_m(z)= [1- (1/I \sum\limits_{i=1}^{I}((\vec{f_{z}^m},i / z )-1) ^2)) )]$. 
\[
%u_m(z)= 1- \frac{1}{D} \sum_{d=1}^{D} \left( \frac{f_{z,d}^m}{z} -1 \right)^2 .
u_m(z)= 1- \frac{1}{D} \sum_{d=1}^{D} \left| \frac{f_{z,d}^m}{z} -1 \right| ^{\alpha}
\]
where the parameter $\alpha$ tunes the relative importance given to small versus large deviations from the expected fractions. 
In Section~\ref{results} we use $\alpha=1$ and $2$.
%u_m(z)= 1- \frac{1}{D} \sum_{d=1}^{D} | \frac{f_{z,d}^m}{z} -1 |^\alpha. where \alpha=1
If $u_m(z)$ is high (close to one), the proportion of top scholars from each discipline is close to $z$, and therefore the impact measure $m$ is able to compensate for discipline bias. This definition of universality satisfies the basic intuition that all metrics are unbiased in the limit $z = 1$. 

Note that $u_m(z) \leq 1$; it can take negative values in contrived biased scenarios. An alternative definition would normalize the deviations from the expected fractions by the variance within each discipline, however this approach would have decreasing universality as $z \rightarrow 1$ due to the increasing variance. This would violate our basic intuition that all metrics are unbiased in the limit $z = 1$.

To eliminate the dependence of the universality assessment on a particular selectivity $z$, we can finally define the \emph{universality index} of $m$: 
\[ \bar{u}_m = \int_0^1 u_m(z) dz. \]
We numerically approximate the integral as:
\[ 
\bar{u}_m \simeq \sum_{q=1}^{99} u_m(q \cdot \Delta z) \, \Delta z, 
\]
where we set $\Delta z=0.01$.
%
%{\sum\limits_{z=0.05}^{0.95} u_{max} (z)} , \]
%

\section{Results}
\label{results}

%changed to twelve from seven and aded arts& humanities in following caption
\begin{figure}[t!]
\centerline{\includegraphics[width=0.6\textwidth]{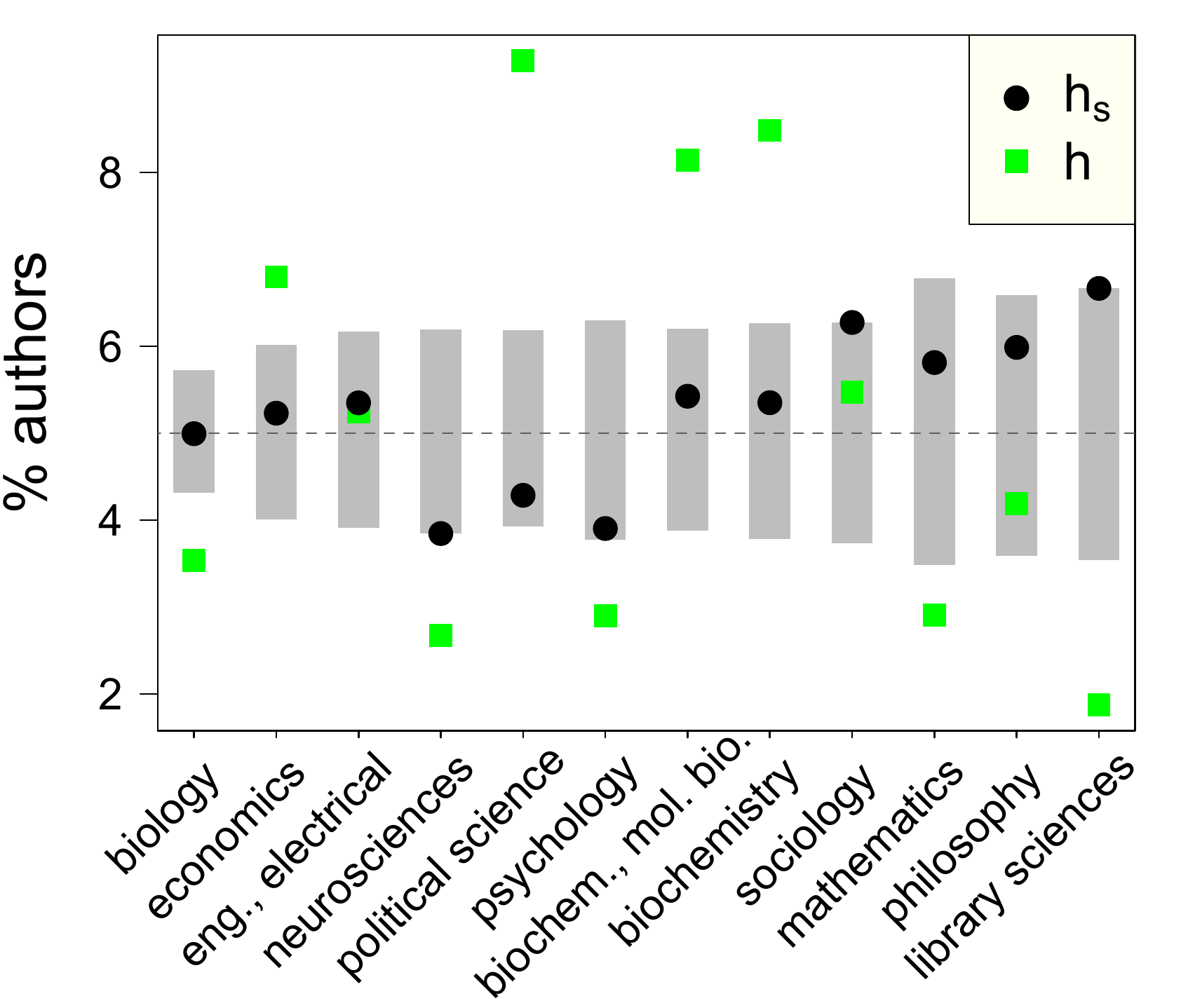}}
\caption{Illustration of discipline bias. The analysis is based on empirical data from the Scholarometer system, which provides discipline annotations for scholars and associated citation material~\cite{Kaur2012}. For legibility we display here only two impact metrics ($h$ and $h_s$) that are used to rank authors in the twelve top JCR disciplines spanning science, social sciences, and arts \& humanities. Across these disciplines, we select the top 5\% of authors according to each metric. We then measure the percentage of authors from this selection that belong to each discipline. The $h$ index favors certain disciplines (e.g., political science) and penalizes others (e.g., library sciences). In this and the following plots, grey areas represent 90\% confidence interval of unbiased samples, as discussed in Section~\ref{theory}.}
\label{fig:combined_b}
\end{figure}

\begin{figure}[t!]
\centerline{\includegraphics[width=0.45\textwidth]{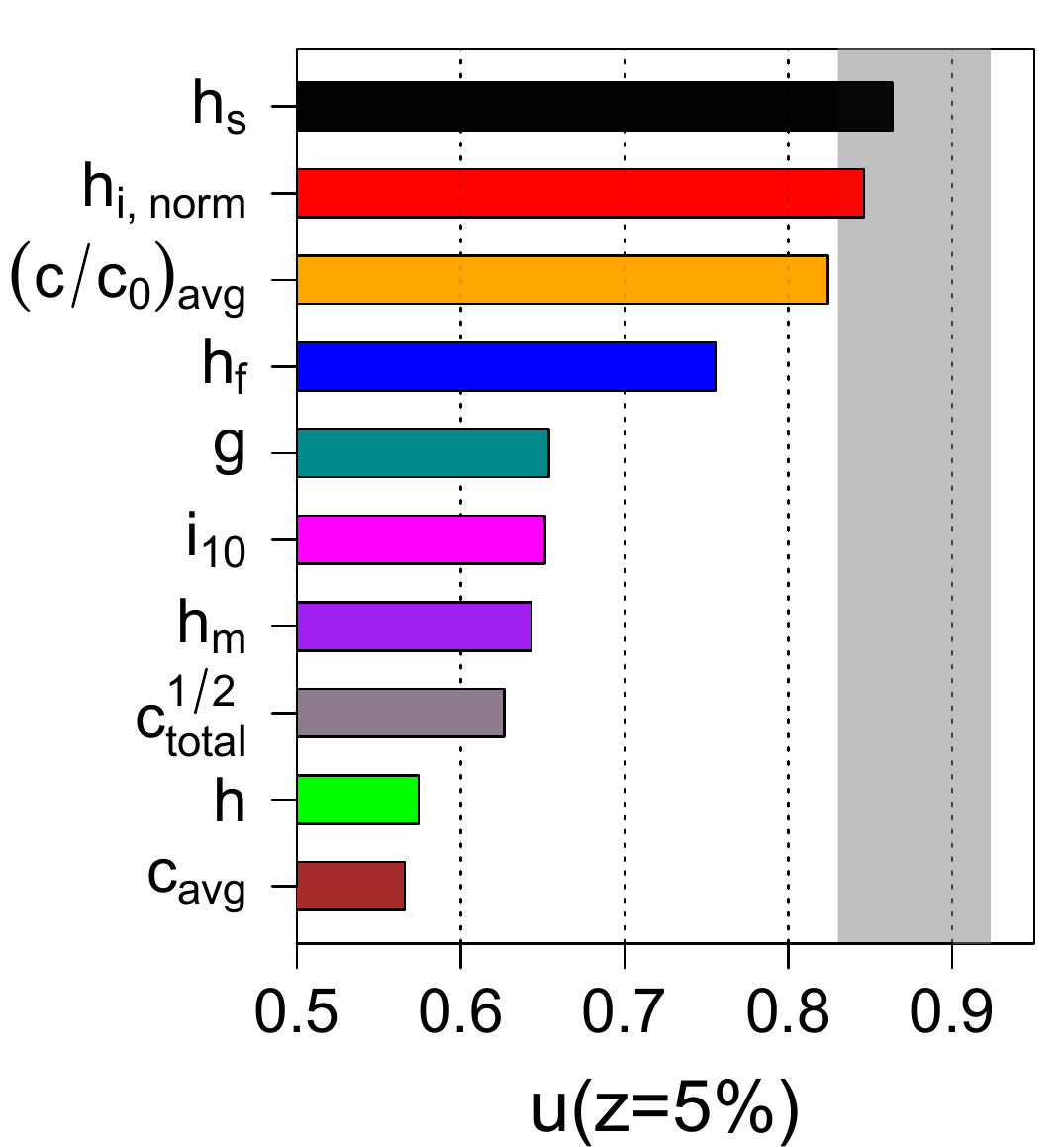}}
\caption{Universality $u(z)$ for ten impact metrics and selectivity $z=5\%$. %The grey area represents the 90\% confidence interval of an unbiased $z=5\%$ sample given by the hypergeometric distribution.
% of $10^3$ simulated samples.
}
\label{fig:box_c}
\end{figure}

\begin{figure}
\centerline{\includegraphics[width=0.95\textwidth]{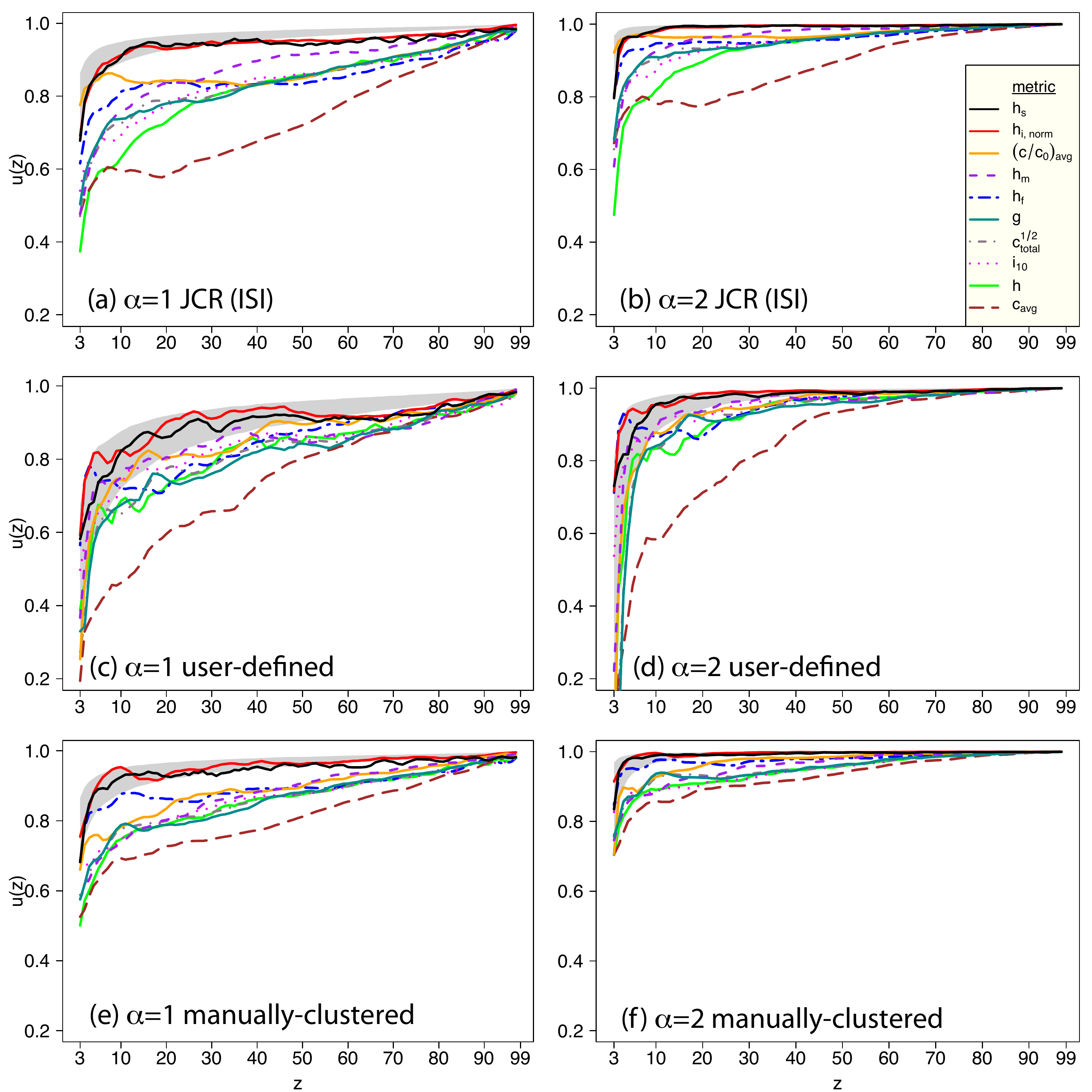}}
\caption{
Universality $u(z)$ as a function of selectivity $z$. Different panels correspond to different categorizations and values for the exponent $\alpha$. Gray areas in the figure display 90\% confidence intervals computed through the null model.
The rankings are not particularly sensitive to different categorizations or exponent values (cf. Table~\ref{summary}).}
\label{fig:All}
\end{figure}

\begin{table}
\caption{Universality index of the ten metrics for different discipline categorizations and exponent values.}
\begin{center}
%\begin{tabular}{rp{1.8in}cp{1.8in}c}
\begin{tabular}{lrrrrrr}
\hline
Metric & \multicolumn{2}{c}{JCR (ISI)} & \multicolumn{2}{c}{User-defined} & \multicolumn{2}{c}{Manual}\\
\hline
&$\alpha=1$&$\alpha=2$&$\alpha=1$&$\alpha=2$&$\alpha=1$&$\alpha=2$\\
\hline
$h_s$ & 0.94 & 0.99 & 0.90 & 0.98 & 0.95 & 0.99\\
$h_{i,norm}$ &  0.94 & 0.99 & 0.90 & 0.98 & 0.95 & 0.99\\
$(c/c_0)_{avg}$ &  0.88 & 0.98 & 0.86 & 0.95 & 0.89 & 0.97\\
$h_m$ &  0.88 & 0.97 & 0.85 & 0.95 & 0.88 & 0.96\\
$h_f$ & 0.86 &  0.96 & 0.85 & 0.95 & 0.90 & 0.98\\
%$(c_avg)/(c_0)_avg$ &   &  &  \\  %change the symbol
$g$ &  0.85 & 0.95 & 0.81 & 0.92 & 0.86 & 0.96\\
$c_{total}^{1/2}$ & 0.85 & 0.95 & 0.82 & 0.93 & 0.87 & 0.96\\
$i_{10}$ &  0.84 & 0.95 & 0.85 & 0.95 & 0.87 & 0.95\\
$h$ & 0.83 & 0.94 & 0.82 & 0.93 & 0.86 & 0.95\\
$c_{avg}$ & 0.74 & 0.89 & 0.74 & 0.85 & 0.81 & 0.93\\
\hline
\end{tabular}
\end{center}
\label{summary}
\end{table}

To illustrate the usefulness of our index, let us analyze the universality for the ten impact metrics described in Section~\ref{impact} across a set of scholarly disciplines. As evident in Fig.~\ref{fig:confidence_ISI}, some metrics are more universal than others. We first consider the disciplines from the Thomson-Reuters JCR classification (see Table~\ref{ISITop10}) for the case $\alpha=1$. 
To better appreciate the different biases, let us focus on just two impact metrics, $h$ and $h_s$ (Fig.~\ref{fig:combined_b}). When we select the top 5\% of all scholars, $h_s$ yields close to 5\% of scholars from each of the considered disciplines, consistently with the null model (grey area); $h$ yields large fluctuations, favoring some disciplines and penalizing others. 

Fig.~\ref{fig:box_c} shows that according to $u(5\%)$, two of the metrics appear to be least biased: Batista's $h_{i,norm}$ and our own $h_s$. These are consistent with the unbiased model at $z=5\%$, while the other metrics are not.

Fig.~\ref{fig:All}(a) shows how the universality of each metric depends on the selectivity $z$. As we select more top scholars, the bias of all metrics decreases; $u(z) \rightarrow 1$ as $z \rightarrow 1$ by definition.
For selectivity $z < 40\%$, the two best metrics display high universality, as illustrated by the overlap of the corresponding curves with the expectation of the null model (grey area). 

%
%For higher $z$, the new crown indicator is surpassed by $h_m$ around $z=25\%$. It is interesting to note that $h_m$ employs normalization by number of co-authors and $h_s$  by field average, while $h_{i,norm}$ employs both.
Table~\ref{summary} reports the values of the universality index $\bar{u}$ integrated across $z$. To evaluate the statistical significance of differences in values of the universality index $\bar{u}$ for different metrics, we need to estimate the fluctuations of this measure. Let us consider the variations in the values of $\bar{u}_{null}$ obtained by simulating the null model for $z\in (0,1)$. Running $10^3$ simulations yields a standard deviation $\sigma_{null}=0.005$. Therefore we do not consider differences in the third decimal digit
statistically significant, and we round $\bar{u}$ values to the second decimal digit. The differences shown are deemed significant. According to this summary, $h_{i,norm}$ and $h_s$ are the  most universal among the impact metrics considered. Their universality indices are statistically equivalent to each other.
% and to the null model ($p=0.14$, see supplementary information). (reason p-value acc to new alpha is not very low 0.003)
The computation of $h_s$ is however much simpler, as it does not require co-author metadata. %The other metrics are not consistent with the null model ($\bar{u} \leq 0.97, p < 0.05$). 

 %added   --or taking into account variance based on discipline size .

%\subsection{Sensitivity analysis}

Next we test the robustness of our findings with respect to several variations of our method: different ways to classify authors into disciplines, different selectivity values, and different exponents in the definition of universality.

\subsection{Sensitivity to discipline definitions}
\label{discipline_sensitivity}

\begin{figure}
\centerline{\includegraphics[width=\textwidth]{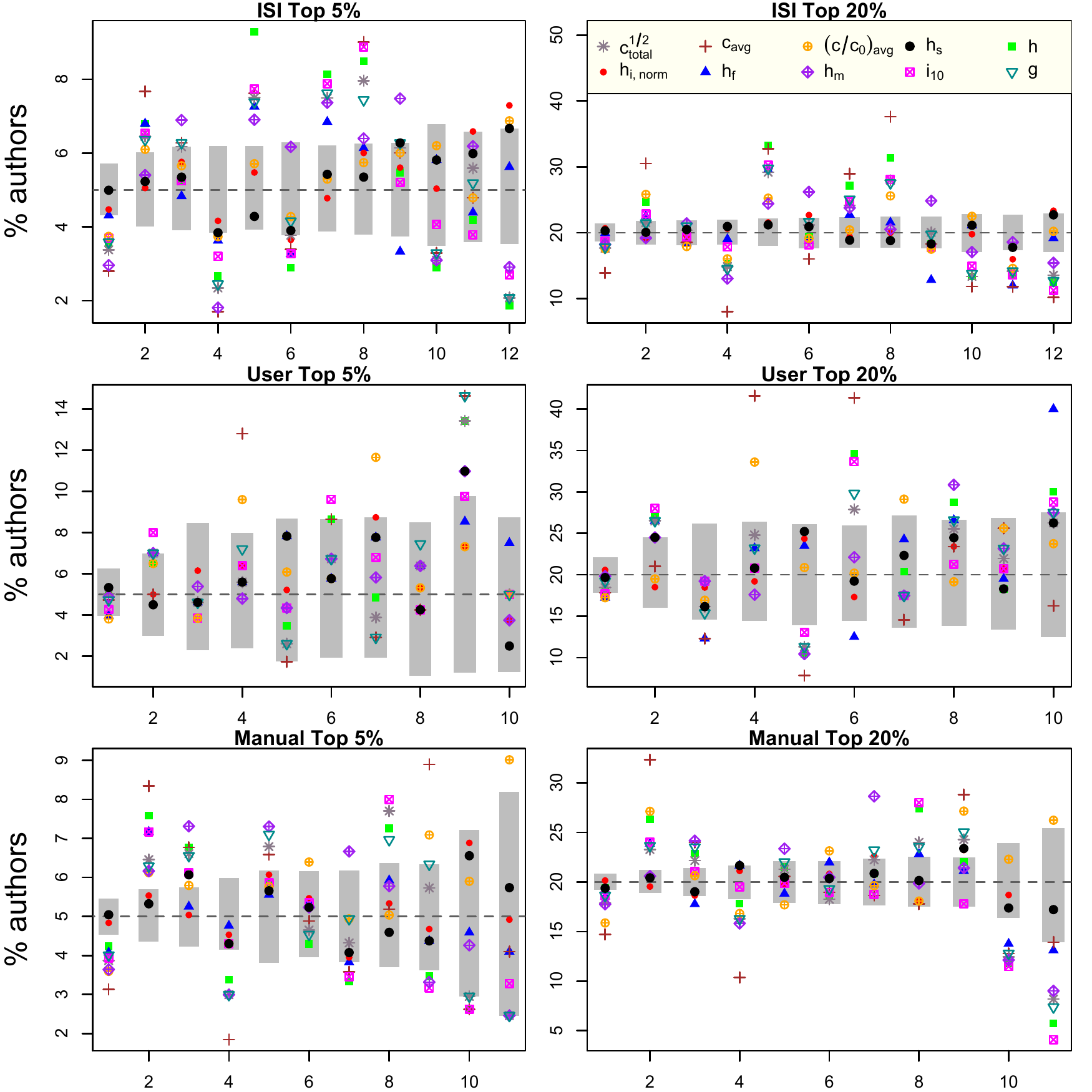}}
\caption{Percentage of authors belonging to different disciplines according to ISI JCR (top), user-defined (middle), and manually-clustered (bottom) disciplines  listed in Section~\ref{disciplines}. The authors are ranked by each metric in the top $z=5\%$ (left) and $20\%$ (right). Gray areas bound the 90\% confidence intervals obtained from the null model.}
\label{fig:confidence_ISI}
\end{figure}

While our definition of universality assumes that authors are associated with disciplines, the results of our analysis are not dependent on the JCR classification.
%Here we extend the analysis in the main paper to the two additional categorization. 
Fig.~\ref{fig:confidence_ISI} extends Fig.~\ref{fig:combined_b} %with more JCR disciplines and
 to the two additional discipline definitions (User and Manual, cf. Section~\ref{disciplines}). The results in all cases are similar.
Fig.~\ref{fig:All} and Table~\ref{summary} show that with a few exceptions, the ranking of impact metrics is consistent across categorizations. In all cases, $h_s$ and $h_{i,norm}$ are the most universal (least biased) metrics.

\subsection{Sensitivity to selectivity $z$}

%explain what we do and show the results
We repeated the analysis of Fig.~\ref{fig:combined_b} for two values of the selectivity parameter $z$. 
%We computed the values for z\% from 1\% to 99\% with interval of 5. We simulated the process explained above with the same number of categories and authors(balls) for $10^3$ times. 
Fig.~\ref{fig:confidence_ISI} shows that for each discipline categorization, the results of the cases $z=0.05$ and $z=0.20$ are statistically similar; the number of times that the measured values are inside the confidence intervals
is not strictly depending on the choice of $z$.

\subsection{Sensitivity to exponent $\alpha$}

Fig.~\ref{fig:All} and the Table~\ref{summary} shows that, with a few exceptions, the ranking of impact metrics is consistent for different exponents $\alpha$. 
%The universality index for all measures is reported in Table~\ref{summary}. 

\section{Conclusion}

%%%%Added%%

While discipline bias is quickly being recognized as a key challenge for objective assessment of impact, it has been problematic until now to evaluate the claims of universality for the multitude of proposed metrics. The index presented here is the first quantitative gauge of universality that can be readily applied to any existing metric. The present analysis points to $h_s$ as an impact metric that is intuitive, easy to compute, and universal. 

The $h_s$ metric does require that the disciplines associated with an author be known,
something that can be a challenge because discipline boundaries are
not sharp \cite{SNIP} and they are continually evolving as new fields
emerge and old ones die \cite{sun2013social}. The
solution we have proposed and implemented in the Scholarometer system
\cite{Kaur2012} is that of crowdsourcing the
discipline annotations of scholars. In this view, annotations are
votes and a scholar is represented as a vector of disciplines. One may
then compute the impact of interdisciplinary scholars with respect to
any relevant discipline, or a combined metric based on the discipline
weights in their vector representations. Further work is needed to verify that such a combination of universal metrics remains universal.

\section*{Acknowledgements}
%check the format
Thanks to Santo Fortunato, Alessandro Flammini, Claudio Castellano, and Yong-Yeol Ahn for helpful feedback on earlier versions of this manuscript. Xiaoling Sun, Diep Thi Hoang, Mohsen JafariAsbagh, and Lino Possamai provided technical support. We acknowledge financial support from the IU School of informatics and Computing, the Lilly Endowment, and NSF (award IIS-0811994) for funding the computing infrastructure that hosts the Scholarometer service. 

%\bibliography{universality}

\end{document}